# Risk Management of AI/ML Software as a Medical Device (SaMD): On ISO 14971 & Related Standards & Guidances

Stephen G. Odaibo,
MD., M.S.(Math), M.S.(Comp. Sci.)

RETINA-AI Health, Inc.

**ABSTRACT**

Safety and efficacy are the paramount objectives of medical device regulation. And in line with the medical ethos of non-maleficence, *first do no harm*, safety is the primary goal of regulation also. As such, risk management is the underlying principle that governs the regulation of medical devices, whether traditional devices or Software as a Medical Device (SaMD). In this article, I review how Risk Management Standard ISO 14971:2019 both connects with and serves as a foundation for the other parts of the Artificial Intelligence (AI)/Machine Learning (ML) SaMD regulatory framework.

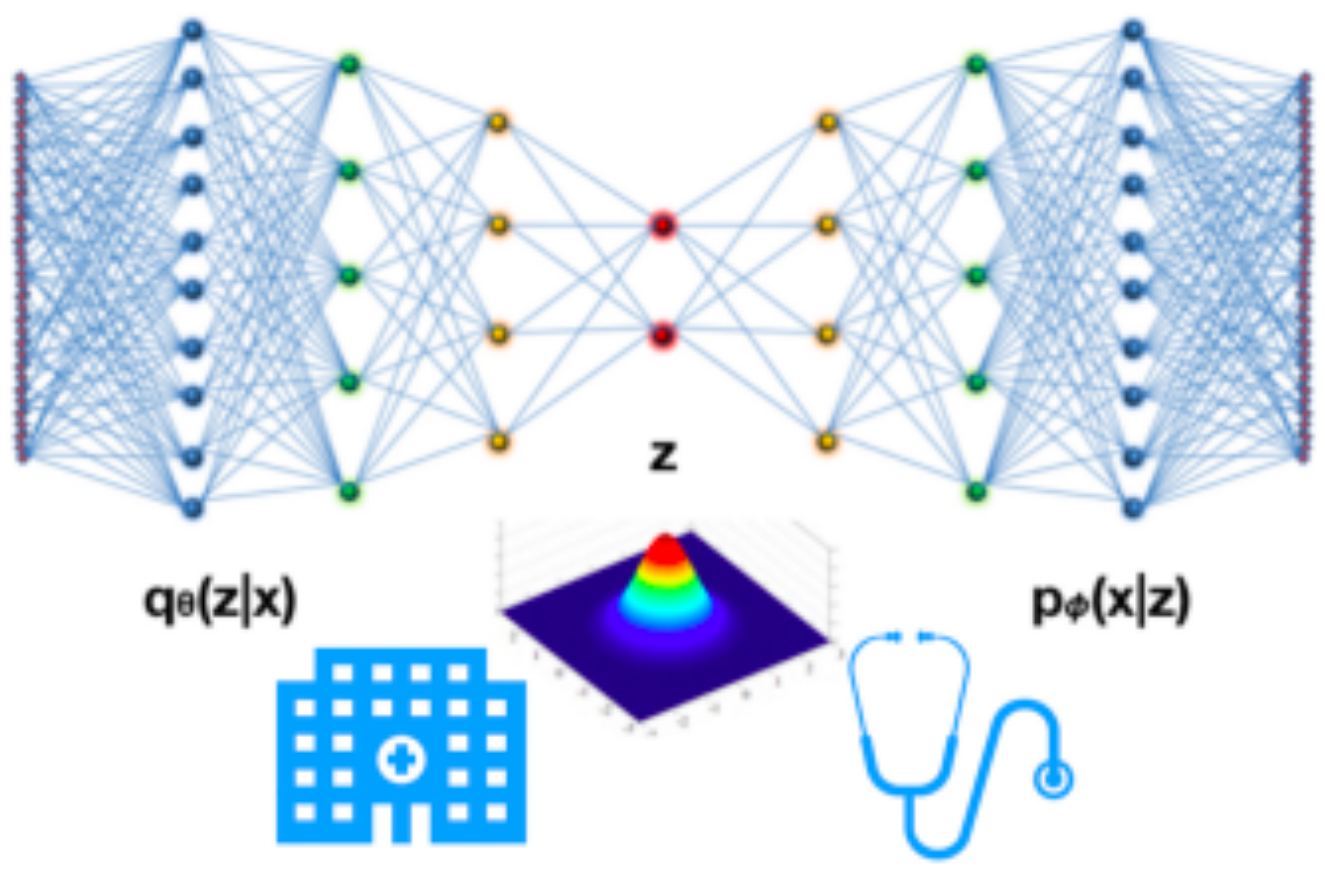

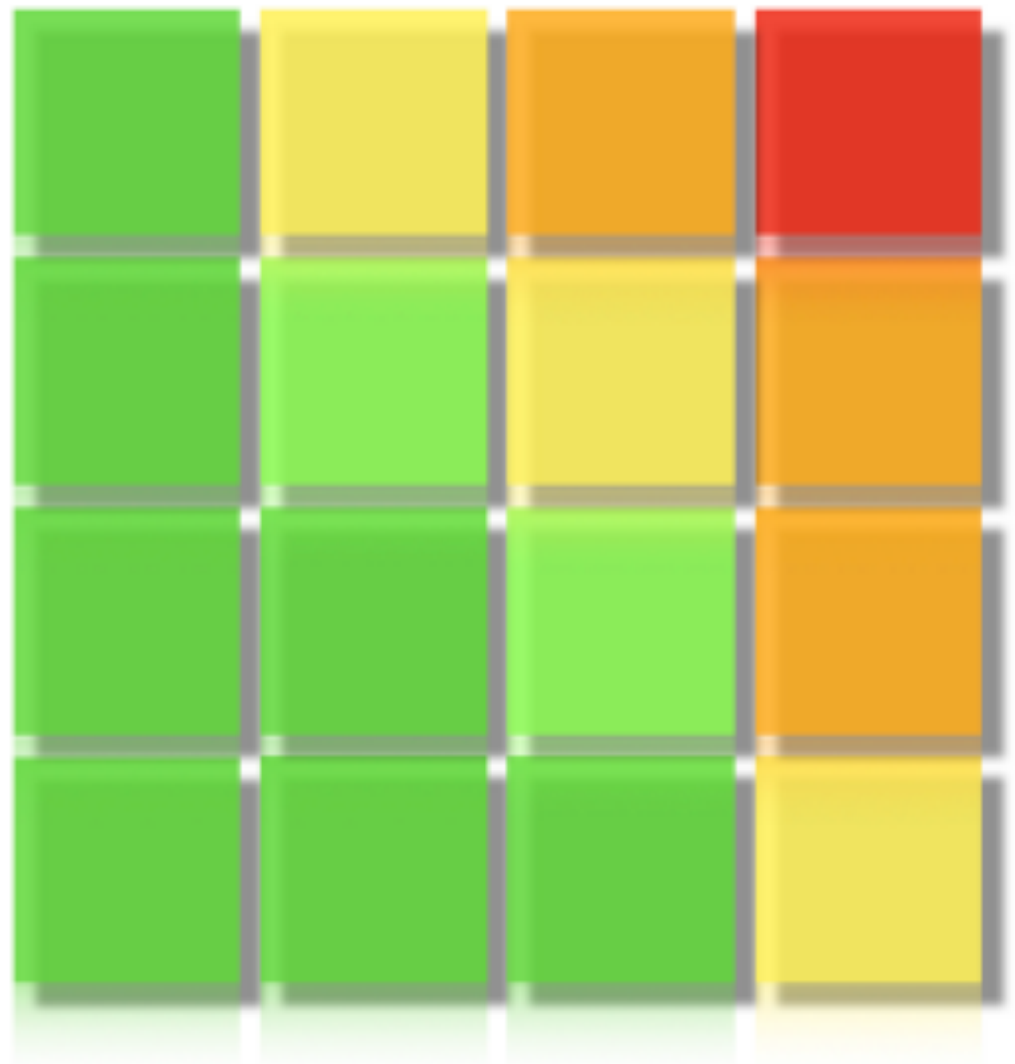

### Introduction:

Furthermore, in some sense the principles of risk management as outlined specifically in ISO 14971 are indeed the underpinning of all medical device regulations and standards. This is true for both traditional medical devices and SaMDs. However, the relatively rapid cycles of development and change of software introduce unique categories of risks and considerations that require special attention. Furthermore, when the SaMDs are Artificial Intelligence (AI) or Machine Learning (ML)-based devices, the distinguishing characteristics of AI/ML-based learning and inference in the pre-market and post-market context require special consideration and handling. This is so as to ensure safe devices and to build trust in the quality and reliability of the service to patients. In this writing, I discuss the state-of-the-art risk management consensus standard, ISO 14971:2019 and its relatives, all in the context of AI/ML SaMD.

Specifically, I will initially (in Section 1) provide an overview of Medical Devices Risk management, ISO 14971 — the main focus of this writing. After which (in Section 2) I will present the relationship of ISO 14971 to Quality Management System (i.e. ISO 13485 and 21 CFR 820) — centering on Design Controls (21 CFR 820.30); then (in Section 3) I will present its relationship to FDA Guidance on requirements for filing 510(k) for SaMD device; then (in Section 4) its relationship to IEC 62304 Software Development Life Cycle (SDLC) for Medical Devices; then (in Section 5) its relationship to FDA Guidance on modifications to SaMD, and finally (in Section 6) I will discuss FDA's Action Plan on AI/ML SaMD.

## Section 1: APPLICATION OF RISK MANAGEMENT TO MEDICAL DEVICES (ISO 14971:2019)

### Overview:

The *risk management process* prescribed by ISO 14971:2019 is as follows. One starts out by determining what potential sources of *harm* (*hazards*) may be associated with a given device (or entity or process). Once hazards are identified, next to explore are the contingencies of *sequences of events* that may lead to *hazardous situations* — situations

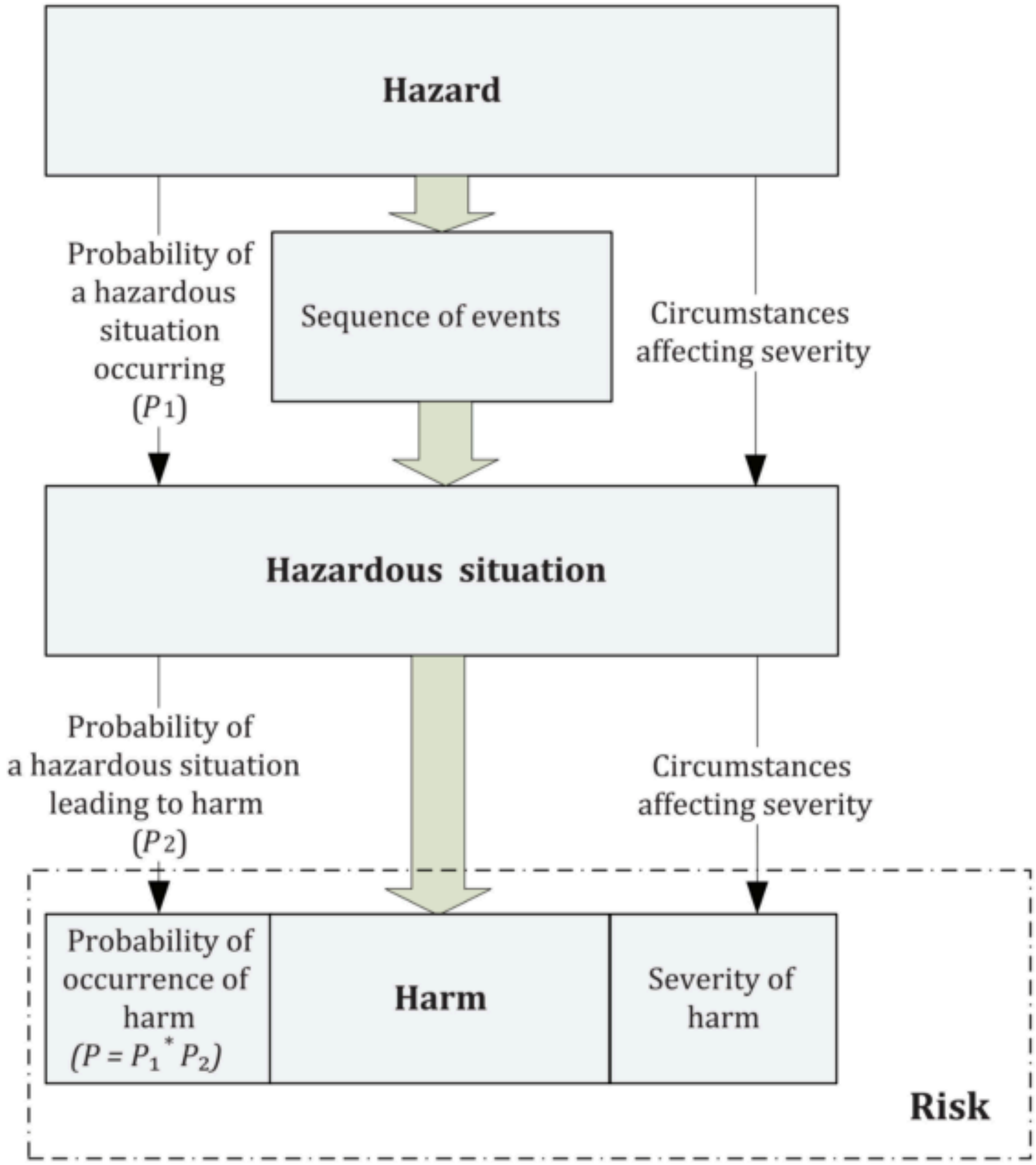

Figure 1: Risk Estimation flow chart from ISO 14971:2019

in which there is exposure to such hazards. Next, is to determine the likelihood of harm and the severity of that harm — the combination of which is the risk; and determining that combination is called *risk estimation* and is illustrated in Figure 1 below

After one has estimated the risk, one consults a policy — the risk acceptability criteria — which dictates whether there is a need to implement *risk control* measures against the said risk. The risk acceptability criteria is pre-determined by the device manufacturer as part of their *risk management plan*. If it is deemed that risk control measures are needed, they are executed after which the *residual risk* is calculated. Risk control measures and the acceptability criteria are iteratively applied to the device till risk is acceptable, or till a *benefit-risk analysis* justifies the device in spite of residual risk. If risk is not reducible to acceptable levels then the device is returned to the design phase for a design modification. The overall residual risk of the device is also determined and evaluated in the same manner as is done for each individual risk. A *risk management review* is then completed at scheduled intervals to assess the execution of the risk management plan and its suitability to manage the risk associated with the medical device. The findings of the risk management review and written into the risk management report. During and after manufacture and release of the device, *production and post-production activities* continue to collect information on the risks of the device and iteratively assess this risk as specified in the risk management plan.

The process of identifying hazards and estimating the associates risks is called *risk analysis*. The process of evaluating the acceptability of estimated risks is *risk evaluation*. At each stage of risk control, the effectiveness of the control measure must be verified. Furthermore, the suitability of the risk management plan must also be verified. And the documentation of all of this — risk analysis, risk assessment, residual risk, risk control, risk control verification, traceability mechanism, delegation of responsibility, evaluation of overall residual risk, risk management review, and production/post-production

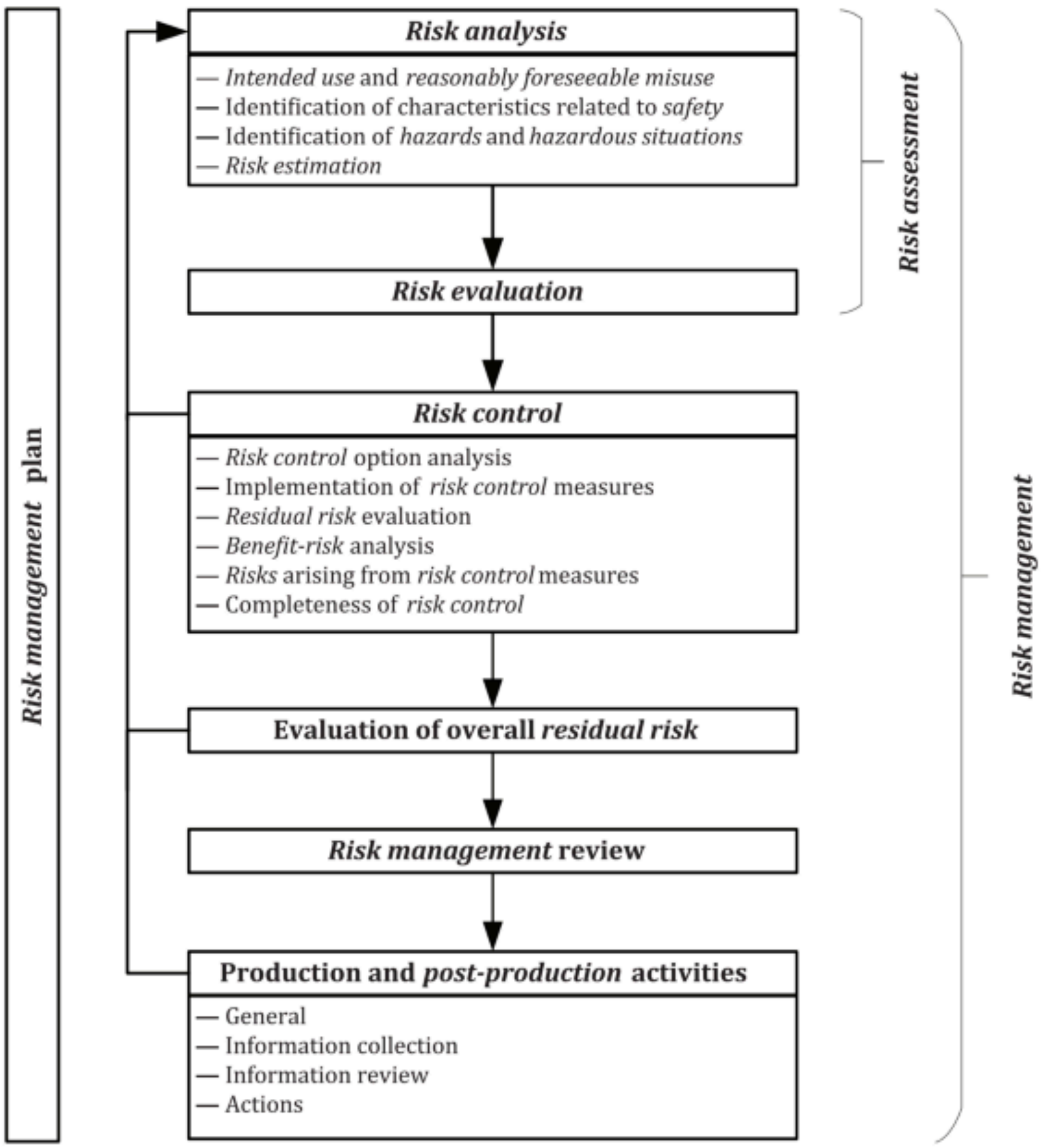

Figure 2: Risk Management Process: From ISO 14971

activities — constitute the Risk Management File (RMF). The overall risk management process is illustrated in Figure 2 below.

### ISO 14971:2019

As mentioned above, ISO 14971:2019 is the consensus standard for risk management in the medical device industry. It consists of the following clauses:

- Clause 1: Scope
- Clause 2: Normative References
- Clause 3: Terms and Definitions
- Clause 4: General Requirements for Risk Management System
- Clause 5: Risk Analysis
- Clause 6: Risk Evaluation
- Clause 7: Risk Control
- Clause 8: Evaluation of Overall Residual Risk
- Clause 9: Risk Management Review
- Clause 10: Production and Post-Production Activities

Figure 3 below illustrates how each Clause maps into the overall risk management process.

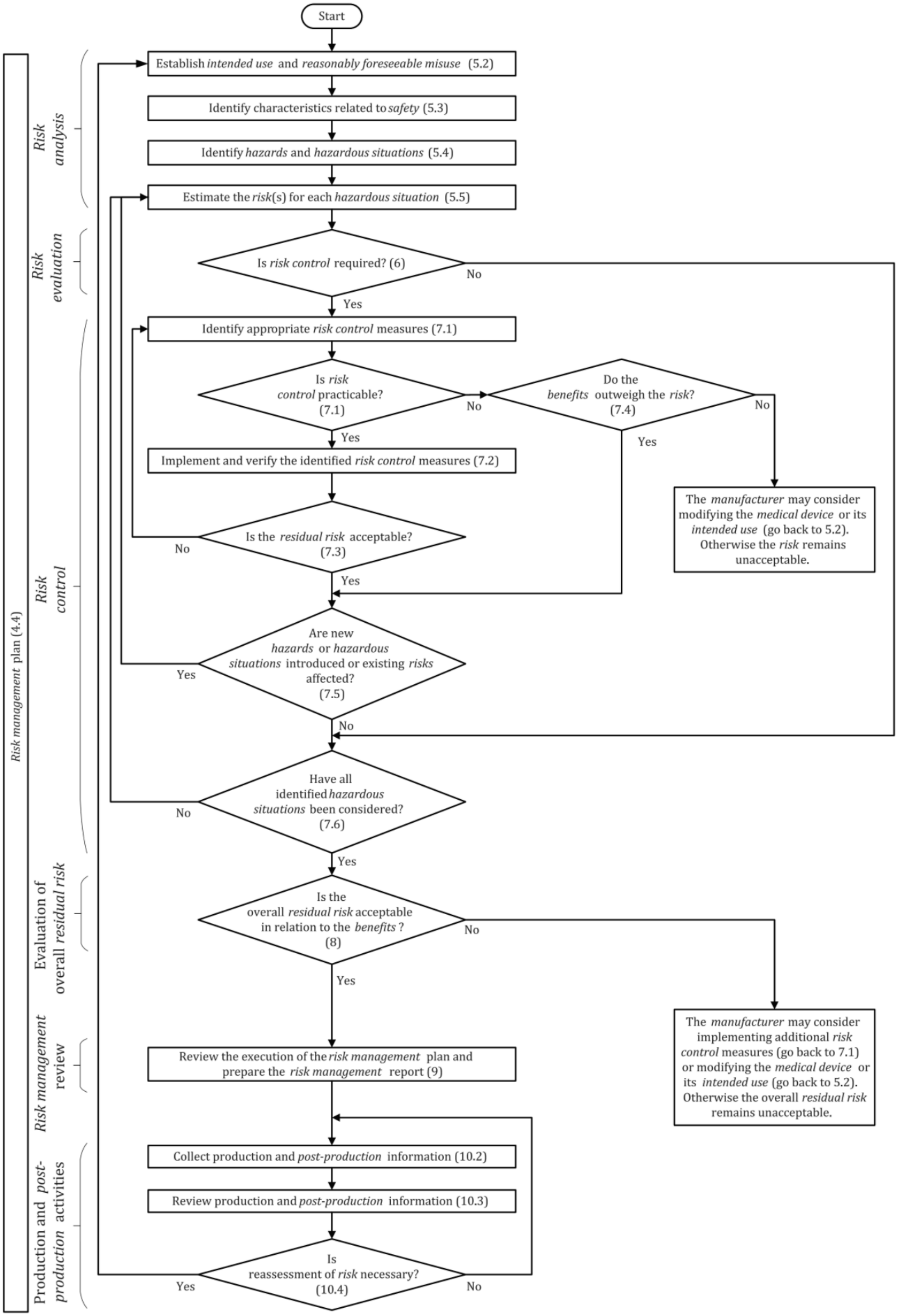

Figure 3: Overall Risk Management Process: From ISO 14971:2019

Some Useful Definitions (from Clause 3 of ISO 14971:2019):

- **Hazard**: Potential source of harm
- **Harm**: Injury or damage to the health of people, or damage to property or the environment **[UPDATED in ISO 14971:2019]**
- **Risk**: Combination of the probability of occurrence of harm and the severity of that harm. [E.g: Risk = (Probability of harm) x (severity of that harm)]
- **Hazardous Situation**: Situation in which people, property, or the environment is exposed to a hazard
- **Risk Analysis**: Determining hazards and estimating their associated risk
- **Risk Estimation**: Determining (or deciding) the probability (or qualitative category) of probability of occurrence of harm and severity of that harm.
- **Risk Assessment**: Risk analysis and risk Evaluation
- **Risk Evaluation**: Comparing the risks with the risk acceptability criteria, to determine if risk control measures are needed.
- **Risk Control**: Process through which decisions are reached and protective measures are implemented for reducing risks to, or maintaining risks within, specified levels [ISO 14971:2000, definition 2.16]

- **Risk Management**: systematic application of management policies, procedures and practices to the tasks of analyzing, evaluating and controlling risk [ISO 14971:2000, definition 2.18]
- **Residual Risk**: Risk remaining after risk control measures have been performed
- **Benefit**: Positive impact or desirable outcome of the use of a medical device (3.10) on the health of an individual, or a positive impact on patient management or public health **[NEW definition to ISO 14971:2019 — See 3.2]**
- **Reasonable Foreseeable Misuse**: Use of a product or system in a way not intended by the manufacturer, but which can result from readily predictable human behavior **[NEW definition to ISO 14971:2019 — See 3.15]**
- **Accompanying Documentation**: Materials accompanying a medical device and containing information for the user or those accountable for the installation, use, maintenance, decommissioning, and disposal of the medical device particularly regarding safe use. **[NEW definition to ISO 14971:2019]**
- **State of the Art**: Developed stage of technical capability at a given time as regards products, processes (3.14) and services, based on the relevant consolidated findings of science, technology, and experience **[NEW definition to ISO 14971:2019 — See 3.28]**

### Historical Pedigree of ISO 14971

- ISO 14971:2000 (First Edition) →
- ISO 14971:2007 (Second Edition) →
- EN-ISO 14971:2009, EN-ISO 14971:2012 →
- ISO/TR 24971:2013 (First Edition) →
- ISO 14971:2019 (Third Edition) →
- ISO/TR 24971:2020 (Second Edition)

### Risk Management System (Clause 4: ISO 14971:2019):

Clause 4 of ISO 14971:2019, titled "Risk Management System," consists of the 4 parts listed below.

- **Clause 4.1**: Risk management process
- **Clause 4.2**: Management responsibilities
- **Clause 4.3:** Competence of personnel
- **Clause 4.4**: Risk management plan
- **Clause 4.5**: Risk management file

### Production and Post-Production Activities (Clause 10: ISO 14971:2019)

This pertains to the activities that occur during and after the device is in production. It aligns with ISO 13485 Clause 8: "Measurement and analysis improvement." And the underlying requirement is that post-production activities cannot be passive, i.e. awaiting bugs in the software to be reported. Instead, the SaMD manufacturer must actively seek feedback on the product via customer surveys, interviews, Corrective Action and Preventive Action (CAPA) reviews, competitor complaints from that product code category; and circling the feedback as a means of internal audit effectiveness. The risk management process and file are not a dormant once and for all output. Instead, must be continuously updated through regularly scheduled reviews (specified in a Company's Risk Management System Standard Operating Procedures) and through information gathering that occurs in the course of production and post-production activities. Clause 10 of ISO 14971:2019 consists of the following 4 parts:

- **General (Clause 10.1)** — general overview of production and post-production activities.
- **Information Collection (Clause 10.2)** — This resourcefully acts on opportunities to obtain information about the risks associated with the device. Specifically, during installation, maintenance, usage, production & post-production activities of the device, a vital trail of information becomes available for analysis and re-evaluation of the device's risk profile. Additionally literature, competitor device performance, and any hints at a general drift of the state-of-the art are also vital sources of information.

- **Information Review (Clause 10.3)** — the primary goal of information review is to keep the risk management file and plan up-to-date and effective in managing the risks associated with the device. It requires that the device manufacturer continually gather information about how the risk profile of the device may be changing due to a number of factors or new information including field complaints, a changing state-of-the-art, changing benefit-risk analysis, etc.
- **Actions (Clause 10.4)** — the information collected and reviewed above is used to update the risk management plan for individual devices as well as used to update the risk management plan itself.

### Some Other Noteworthy Points on ISO 14971

- ISO/TR 24971 is a supporting technical report (TR)to help with implementation of the requirements (numbered clauses) of ISO 14971.
- Especially relevant to SaMD is “Risks Related to Cyber/Data Security” (Annex F of ISO 24971:2020).
- Similarly to ISO/TR 24971, the contents of the Annexes of the ISO 14971 are guidances and not requirements. For instance in the latest release ISO 14971:2019 and ISO/TR 24971:2020, some content such as *risk related to cybersecurity/data* has been moved from 14971:2007 to Annex F of 24971:2020.
- FDA has accepted ISO 14971:2019 for risk management, and will cease to recognize ISO 14971:2007 in December 2022.
- ISO 14971:2019 does not pertain to business risk management, ISO 31000

### Risks Related to Cyber/Data Security (ISO/TR 24971 Annex F)

Security risk management follows the same risk management process of ISO 14971:2019 described above. **Some Definition of Terms, Annex F.2**:

- **Security**: a condition that results from the establishment and maintenance of protective measures that ensure a state of inviolability from hostile acts or influences (see 3.13 in IEC Guide 120:2018[4]), where hostile acts or influences could be intentional or unintentional.
- **Threat**: potential for violation of security, which exists when there is a circumstance, capability, action, or event that could breach security and cause harm (see 3.16 in IEC Guide 120:2018[4]). Threat corresponds to an event or a sequence of events that can exploit a vulnerability leading to a hazardous situation (see 3.5 in ISO 14971:2019).
- **Vulnerability**: flaw or weakness in a system's design, implementation, or operation and management that could be exploited to violate the system's security policy (see 3.18 in IEC Guide 120:2018[4]). Vulnerability can be seen as a type of event or circumstance (see Table C.2 in ISO 14971:2019).
- **Confidentiality**: property that information is not made available or disclosed to unauthorized individuals, entities, or processes (see 3.6 in IEC Guide 120:2018[4]).
- **Integrity**: property of accuracy and completeness (see 3.9 in IEC Guide 120:2018[4]).
- **Availability**: property of being accessible and usable upon demand by an authorized entity (see 3.5 in IEC Guide 120:2018[4]).

**A Mischaracterization in Table F.1 in ISO/TR 24971:2020 Annex F:**

Table F.1 of ISO/TR 24971:2020 refers to "loss of data confidentiality" as a hazard but does not refer to it as a harm. I do not agree with this characterization, as *loss of data confidentiality* is a harm to the patient due to privacy violation per HIPAA. Indeed it can also be deemed a hazard in the sense that every harm may potentially lead to more harm, but this is superfluous and not informative. The circumstance under which a loss of confidentiality can be correctly deemed a hazard and not a harm is under ISO 14971 editions which pre-dated ISO 14971:2019 (i.e. 2000 and 2007 editions). In those previous editions the definition of harm required that the damage done be physical, as such loss of confidentiality may potentially be deemed as not meeting the definition of a harm. This condition further requires the assumption that bits and bytes are not physical. In summary, this categorization in Table F.1 is erroneous considering that the technical report ISO/TR 24971:2020 was released after the definition of a harm was already updated in 2019 to include the word physical; and considering that the ISO/TR 24971:2020's explicit purpose is to support and help in the implementation of ISO 14971:2019.

## SECTION 2: RISK MANAGEMENT of AI/ML SaMD in a QUALITY MANAGEMENT SYSTEM (QMS)

As one conceptualizes, designs, and develops an AI/ML SaMD, hazards exists at every stage of the pipeline. Within the context of a QMS to ensure a quality product, one can and should apply the principles of risk management to the process of quality management. The International Standard for Quality Management System is ISO 13485:2016. In the U.S. the regulation requiring a Quality Management System is 21 CFR Part 820. The FDA has expressed interest in moving towards ISO 13485. Specifically, the FDA stated that: "*FDA intends to harmonize and modernize the Quality System regulation for medical devices. The revisions will supplant the existing requirements with the specifications of an international consensus standard for medical device manufacture, ISO 13485:2016. The revisions are intended to reduce compliance and record keeping burdens on device manufacturers by harmonizing domestic and international requirements. The revisions will also modernize the regulation.*" ISO 13485:2016 consists of the following clauses:

- Clause 1: Scope
- Clause 2: Normative References
- Clause 3: Terms and Definitions
- Clause 4: Quality Management System
- Clause 5: Management Responsibility

- Clause 6: Resource Management
- Clause 7: Product Realization
- Clause 8: Measurement, Analysis, and Improvement

A critical step in the process is *Design Controls* — specified in 21 CFR Part 820.30 which maps to *Design and Development* (7.3) of *Product Realization* (Clause 7). Figure 4 below illustrates Design Controls.

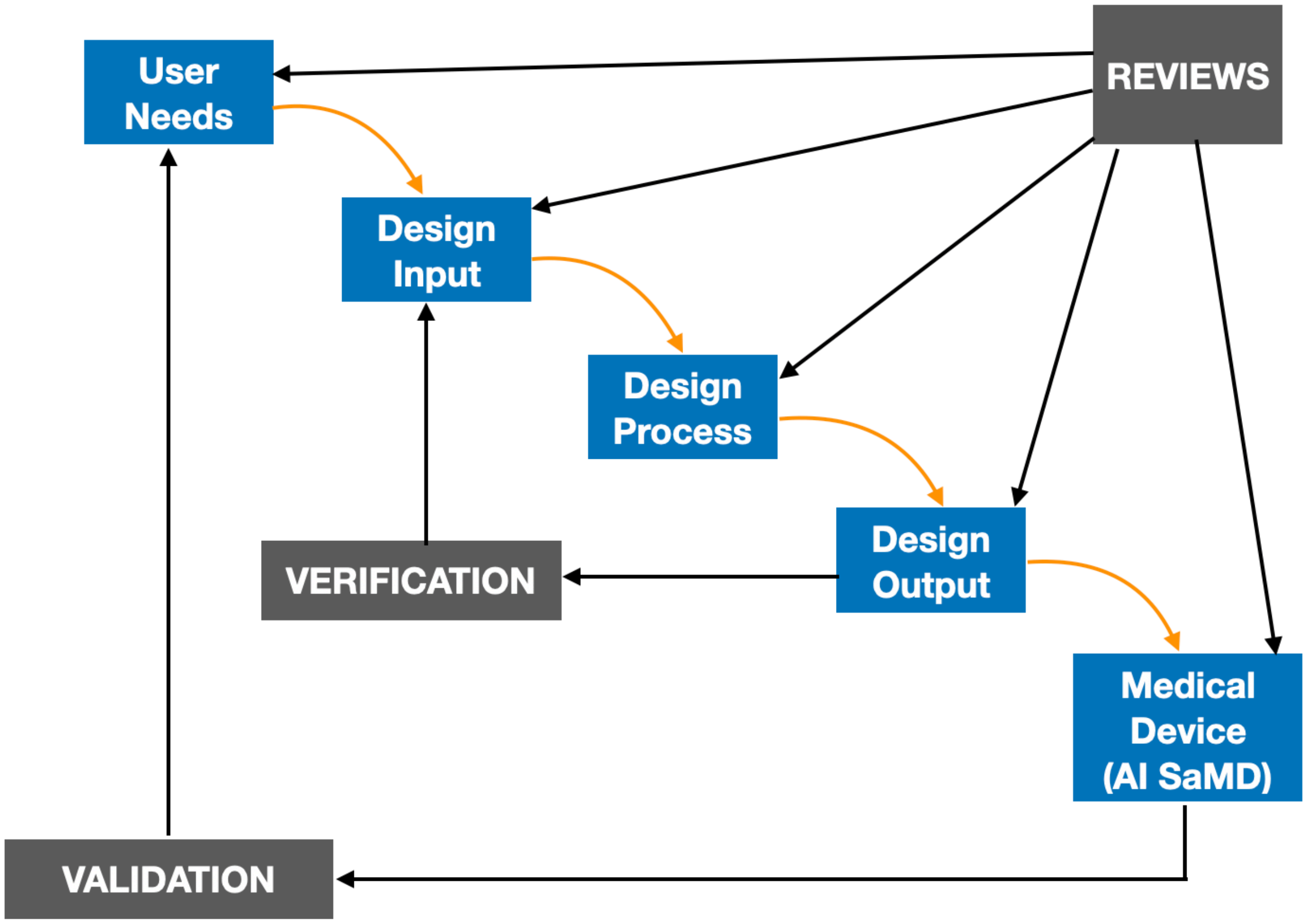

Figure 4: Design Controls

Each need to control each identified risk per risk analysis step from ISO 14971:2019 can be cast or viewed as a design input; and the corresponding risk control measure can be cast or viewed as the corresponding design process. Conversely, software requirement specification resulting from a user need can be cast or viewed as a need to control a corresponding risk. In other words, each design input is a particular need to control some risk, and the corresponding design process is the risk control measure. And the verification requirement of Design controls maps to the verification requirement of risk management. This equivalence applies not only to Product Realization (Clause 7), but to all aspects of the QMS. Since hazards can be associated with aspect of the QMS, the tools and techniques of risk management are applicable. Figure 5 below illustrates the International Medical Device Regulators Forum (IMDRF) Guidance on application of *Risk Management* (Clause 4 of ISO 14971:2016) to *Design and Development* (Clause 7.3 of ISO 13485:2016). While Figure 6 illustrates the IMDRF Guidance on integrating Corrective Action and Preventive Action (CAPA) into Risk management.

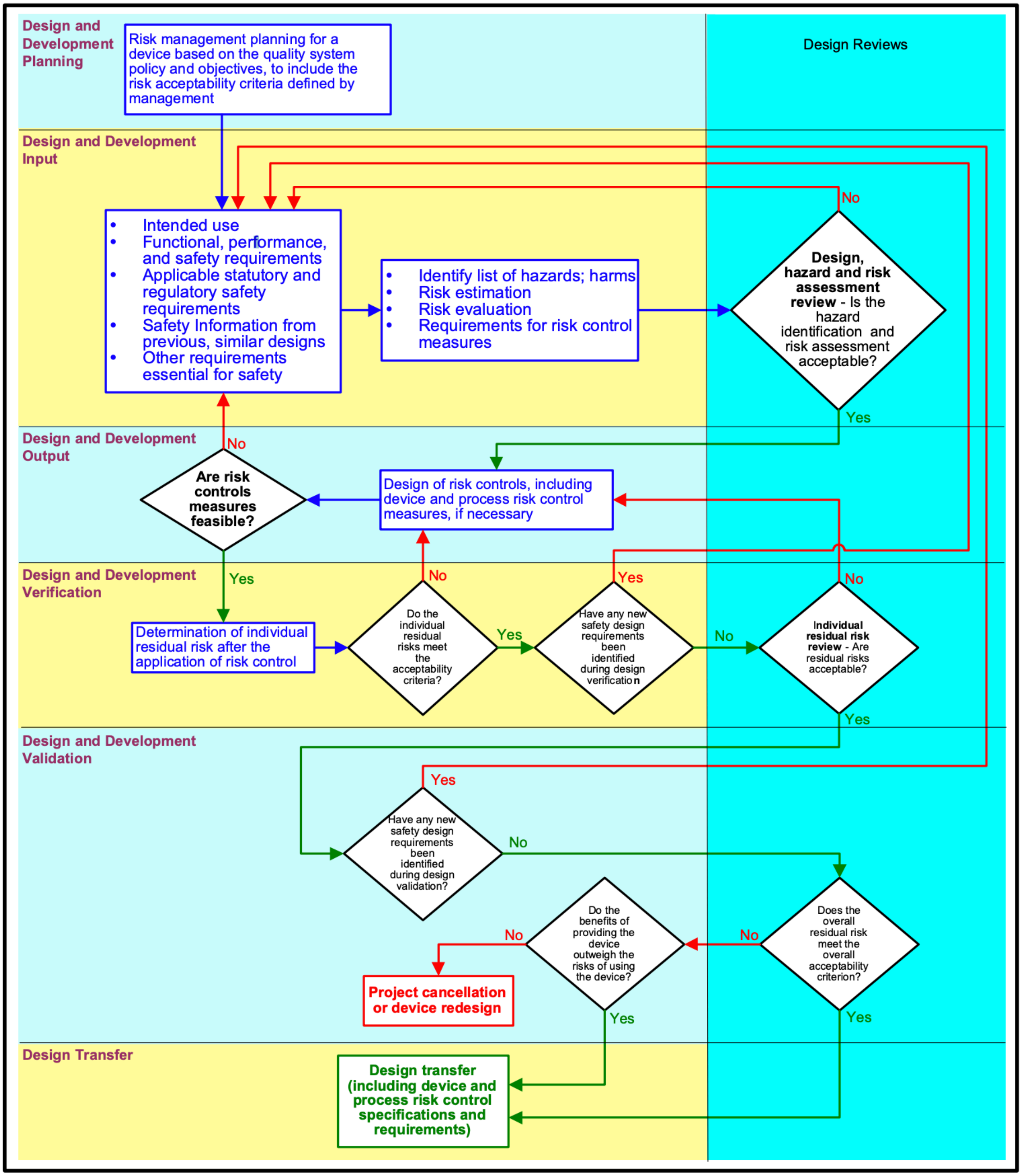

Figure 5: IMDRF Guidance on Application of Risk Management Principles to Design and Development

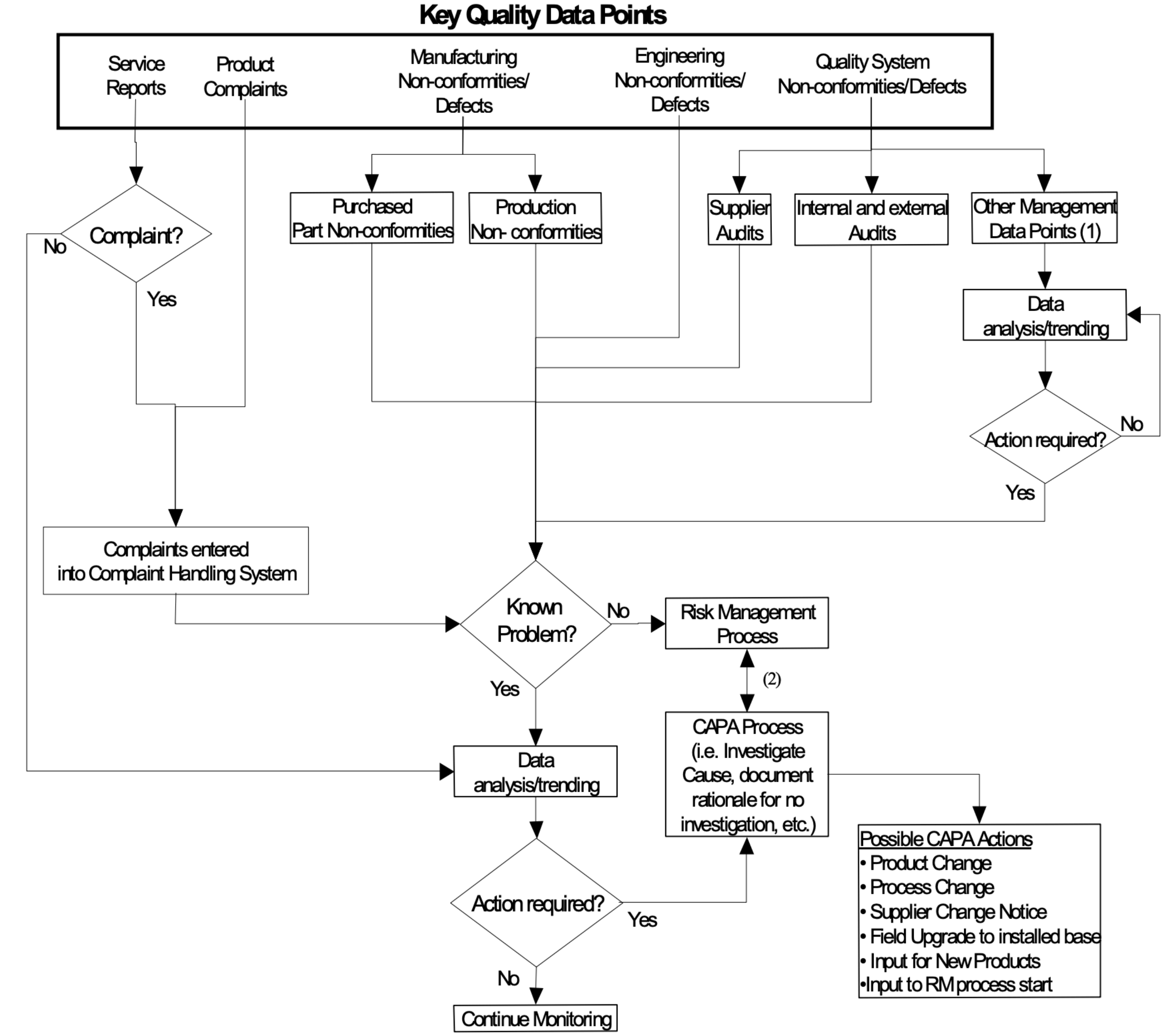

(1) Such as Finished Goods Returned, Credit restock
(2) The relationship will depend upon the output of the investigation. This process can be iterative

Figure 6: IMDRF Guidance on Integrating Corrective Action and Preventive Action (CAPA) into Risk Management

## SECTION 3: RISK MANAGEMENT REQUIRED in FDA 510(k) APPLICATIONS for SaMD

The FDA Guidance on what to submit with a 510(k) titled, *Guidance for the Content of Premarket Submissions for Software Contained in Medical Devices*, includes the following:

- Level of Concern Document
- Software Description
- **Device Hazard Analysis**
- Software Requirement Specifications (SRS)
- Architecture Design Chart
- Software Design Specifications (SDS)
- Traceability Analysis
- Software Development Environment Description
- Verification and Validation (V&V) Documentation
- Revision Level History
- Unresolved Anomalies (Bugs or Defects)

Note the *Device Hazard Analysis* (#3 above) is included. Of note, the term *Device Hazard Analysis* used in the guidance is different from the ISO 14971:2019 term of *Risk Analysis.* And they do not refer to the same thing. Regarding the Device Hazard Analysis, the FDA Guidance recommends a tabular format with a line for each hazard that it should include:

- identification of the hazardous event
- severity of the hazard
- cause(s) of the hazard
- method of control (e.g alarm or hardware design)
- corrective measures taken, including an explanation of the aspects of the device design/requirements that eliminate, reduce, or warn of a hazardous event, and
- verification that the method of control was implemented correctly.

Overall one can say that in ISO 14971:2019 parlance, the FDA term “Device Hazard Analysis” approximately implies: Risk Analysis + Risk Evaluation + Risk Control + Verification of Risk control + Traceability.

## SECTION 4: RISK MANAGEMENT in IEC 62304 SOFTWARE DEVELOPMENT LIFE CYCLE

Software development life cycle (SDLC) is guided by standard IEC 62304 which consists of the following parts:

- Clause 1: Scope
- Clause 2: Normative References
- Clause 3: Terms and Definitions
- Clause 4: General requirements (including 4.2: **Risk Management**)
- Clause 5: Software Development Process
- Clause 6: Software Maintenance Process
- Clause 7: **Software Risk Management Process**
- Clause 8: Software Configuration Management Process
- Clause 9: Software Problem Resolution Process

Clause 4.2 and Clause 7 deal explicitly with risk management as outlined in ISO 14971:2019. Clause 4.2 explicitly states that “The MANUFACTURER shall apply a RISK

MANAGEMENT PROCESS complying with ISO 14971." And Clause 7 outlines step-by-step the process of 14971 as pertains specifically to medical device software. Notably, all aspects of the IEC 62304 standard are in place to minimize the risk associated with the relevant stage of the total life cycle of the SaMD.

### SECTION 5: FDA's RISK-BASED GUIDANCE to SaMD MODIFICATIONS

Unlike traditional medical devices that are primarily hardware and largely static, Software as a Medical Device is protean and ever changing. It is continuously being monitored, patched, updated. The FDA is fully aware of this reality and has provided guidance titled, *Deciding When to Submit a 510(k) for a Software Change to an Existing Device — published October 25th 2017.* The guidance relies on the assumption of a rigorous Quality Management System, and against this backdrop it takes a *least burdensome approach* to managing software changes. For any changes made with intent to "significantly affect the safety or effectiveness of a device … submission of a new 510(k) is likely required." Otherwise one should leverage the principles of risk management and proceed as outlined in the flow chart Figure 7 below:

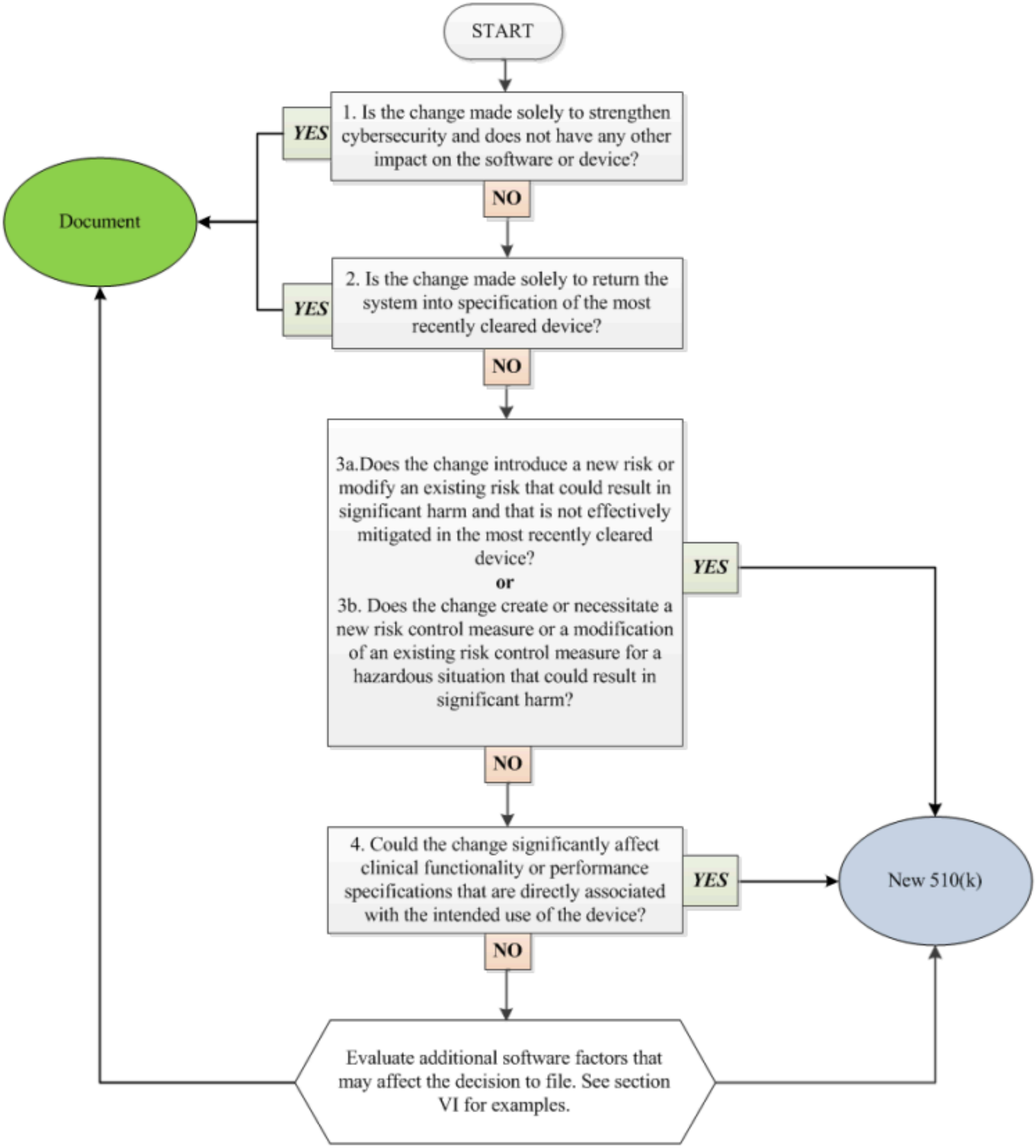

Figure 7: FDA Guidance on “When to Submit a New 510(k) for a Software Change to an Existing Medical Device

## SECTION 6: FDA's ACTION PLAN on AI/ML SaMD

Due to the special character or capability of AI/ML Software to learn and improve with increasing exposure to real world data, the FDA recognized this as a distinct feature requiring some dialogue. As such an Action Plan was published in January 12th 2021 in response to the stakeholder feedback to the April 2019 discussion paper and request for information, *Proposed Regulatory Framework for Modifications to Artificial Intelligence/ Machine Learning-Based Software as a Medical Device*. I contributed feedback to that request for information via Alliance for Artificial Intelligence in Healthcare expert working group. The resulting Action Plan expresses the FDA's thinking and intended actions regarding AI/ML SaMD specific regulation.

**Predetermined Change Control Plan:** consists of:

- SaMD Pre-Specifications (SPS)
- Algorithm Change Protocol (ACP)

The SPS pre-specifies the changes the device manufacturer anticipates making to the device, while the ACP is the algorithm via which such change will be achieved. In other words, the SPS is the "what" while the ACP is the "how." Under-guarding all of this is the principle of risk management as outlined in ISO 14971:2019. Making controlled pre-specified changes, verification and validation the effects of those changes, and iteratively assessing the impact on the device risk profile at every stage of the process. The FDA's proposed regulatory pathway for modifications to AI/ML-based SaMDs is shown in Figure 8 below.

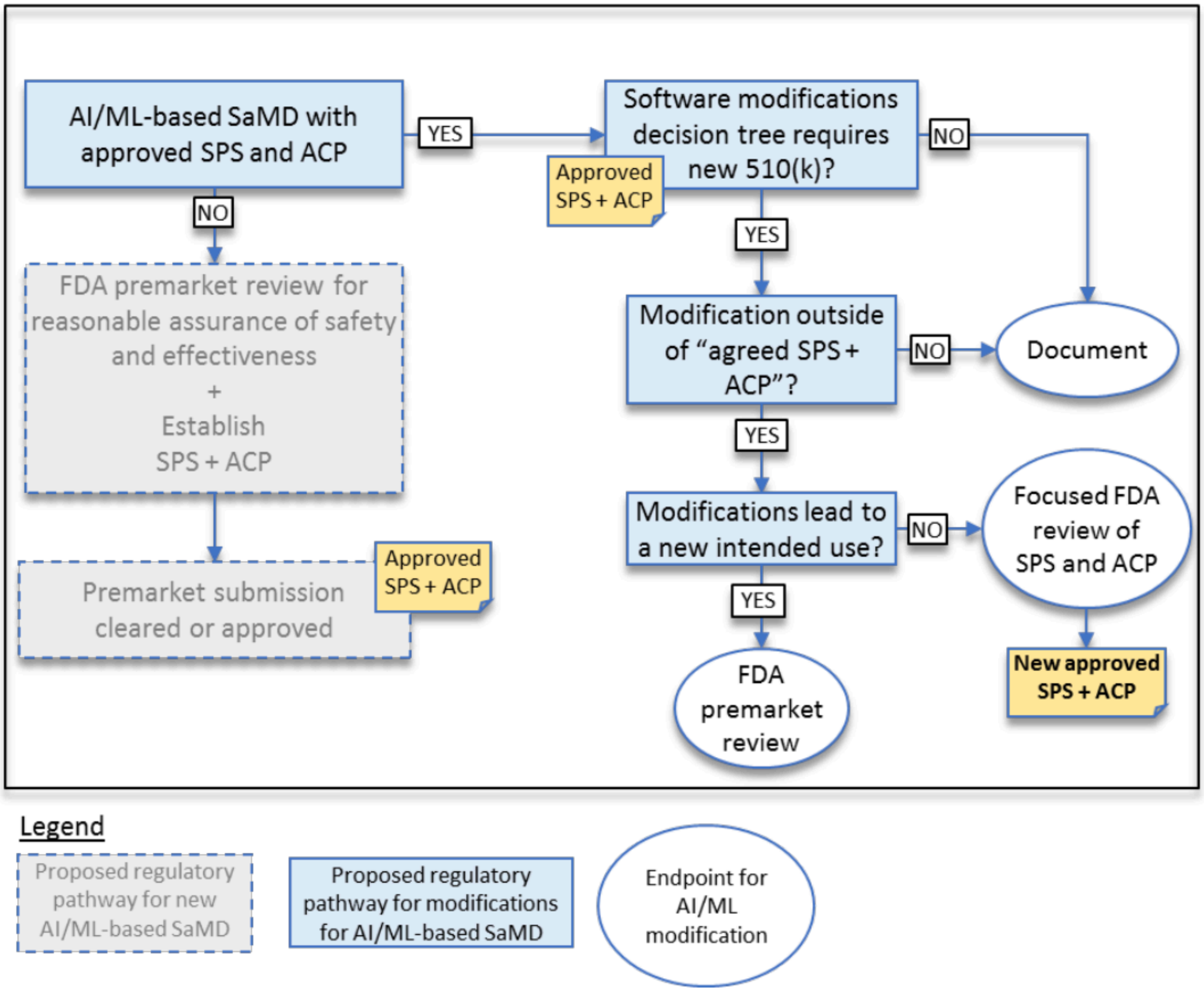

Figure 8: FDA's proposed regulatory pathway for modifications to AI/ML-based SaMDs

The FDA intends to publish a draft Guidance in 2021 to include clarification on the timeframe and submission process of a "Focused Review" and specific guidelines for the SPS and ACP. The Focused Review can be requested when the intended changes are not covered by the agreed SPS+ACP, but do not lead to a new Intended Use.

## CONCLUSION

The motif of Design Controls (Design Input → Design Process → Design Output) can serve as a basis for expressing essentially any aspect of a Quality Management System or objective such as a need for risk control. This simple transformation rule enables full integration of the regulatory standards and frameworks pertinent to AI/ML SaMD, and an expression of AI/ML SaMD regulatory compliance as safety maximization (risk minimization) in a manner that enables production of safe high quality medical devices.

***AUTHOR BIO:*** *Dr. Stephen G. Odaibo is CEO & Founder of RETINA-AI Health, Inc. He is a Physician, Retina Specialist, Mathematician, Computer Scientist, and Full Stack AI Engineer. In 2021 he was issued a U.S. Patent for inventing an AI system that automatically detects diseases from ophthalmic images. In 2017 he received UAB College of Arts & Sciences' highest honor, the Distinguished Alumni Achievement Award. And in 2005 he won the Barrie Hurwitz Award for Excellence in Neurology at Duke Univ School of Medicine where he topped the class in Neurology and in Pediatrics. He is author of the books "Quantum Mechanics & The MRI Machine" and "The Form of Finite Groups: A Course on Finite Group Theory." Dr. Odaibo Chaired the "Artificial Intelligence & Tech in Medicine Symposium" at the 2019 National Medical Association Meeting. Through RETINA-AI, he and his exceptionally talented team are building AI solutions to address the world's most pressing healthcare problems. He resides in Houston Texas with his family.*